 \documentclass[12pt,preprint]{aastex}

\usepackage{epsf}

\newcommand{\mjb}{mJy~beam$^{-1}$}
\newcommand\jb{Jy~beam$^{-1}$}
\newcommand\kms{km~s$^{-1}$}

\shorttitle{OH (1720 MHz) Masers in SNR W44}
\shortauthors{Hoffman et al.}

\begin{document}

\title{The OH (1720 MHz) Supernova Remnant Masers in W44: \\ MERLIN and VLBA Polarization Observations}

\author{Ian M.\ Hoffman}
\affil{National Radio Astronomy Observatory, P.\ O.\ Box O, Socorro, NM, USA 87801}
\affil{Department of Physics and Astronomy, University of New Mexico, Albuquerque, NM, USA 87131}
\email{ihoffman@nrao.edu}

\author{W.\ M.\ Goss}
\affil{National Radio Astronomy Observatory, P.\ O.\ Box O, Socorro, NM, USA 87801}

\author{C.\ L.\ Brogan}
\affil{Institute for Astronomy, 640 North A`ohoku Place, Hilo, HI, USA 96720}

\and

\author{M.\ J Claussen}
\affil{National Radio Astronomy Observatory, P.\ O.\ Box O, Socorro, NM, USA 87801}

\begin{abstract}
Full-polarization MERLIN and VLBA observations of the 1720-MHz maser emission from the OH molecule in the supernova remnant W44 are presented.
Zeeman splitting has been directly resolved between right- and left-circularly polarized spectra indicating magnetic field strengths of $|\mathbf{B}| \approx 0.50$~mG.
The position angle of the linear polarization is in good agreement with the orientation of the shocked gas at the supernova remnant/molecular cloud interface.
Based on MERLIN data having a resolution of 200 milliarcseconds (mas) and VLBA data with a resolution of 15~mas, the masers are measured to have deconvolved angular sizes of 50 to 350~mas (150 to 1000~AU) with compact cores 20~mas (60~AU) in size, consistent with theoretical expectation and previous observations.
\end{abstract}


\section{Introduction}

Spectral-line emission from the OH (1720 MHz) satellite transition $({^2}{\Pi}_{3/2},J={3\over2},F=2{\rightarrow}1)$ in supernova remnants (SNR's) was first observed by Goss (1968) toward W28 and W44.
The SNR OH emission in the 1720~MHz line is non-thermal and is accompanied by absorption in the 1612-, 1665-, and 1667-MHz ground state hyperfine transitions of OH (Goss \& Robinson 1968; Ball \& Staelin 1968; Turner 1969; Robinson, Goss, \& Manchester 1970; Goss, Caswell, \& Robinson 1971).
Beginning with observations of W28 by Frail, Goss, and Slysh (1994), subsequent studies and surveys (Yusef-Zadeh et al.\ 2000; Yusef-Zadeh et al.\ 1999b; Koralesky et al.\ 1998; Green et al.\ 1997; Frail et al.\ 1996; Yusef-Zadeh et al.\ 1996) have observed $\sim 200$ Galactic supernova remnants and found 22 remnants with associated OH (1720~MHz) maser emission.

Elitzur (1976) suggests that the level inversion of these 1720~MHz SNR masers is due to collisional excitation.
The physical conditions suggested for the collisional pump ($n \approx 10^5\,{\rm cm}^{-3}$, $T \approx 90$~K, $N_{\rm OH} \sim 10^{16-17}\,{\rm cm}^{-2}$) are observed to occur where a C-type (non-dissociative) shocked interface between the SNR and an adjacent molecular cloud is moving transversely across the sky ({\it e.g}.\ Elitzur 1976; Wardle 1999; Lockett, Gauthier, \& Elitzur 1999; see also Frail et al.\ 1996).
The temperature, density, and OH column of the OH (1720~MHz) collisional pump are well established from observation ({\it e.g}.\ Claussen et al.\ 1997, hereafter C97; Hoffman et al.\ 2003a, hereafter H03).  
Furthermore, both Claussen et al.\ (2002) and H03 find the intrinsic sizes of the masers to be consistent with predictions based on collisional pump theory (${\sim}10^{15}$~cm, Lockett et al.\ 1999).
Also, observations of the shocked emission from CO ($J = 3 \rightarrow 2$) near the OH masers in W28 and W44 by Frail \& Mitchell (1998) and Arikawa et al.\ (1999) indicate that the masers originate in the post-shock gas, consistent with the collisional-pump model (see also Reynoso \& Mangum 2000; Dubner et al.\ 2004; Lazendic et al.\ 2004).

OH (1720~MHz) SNR masers exhibit relatively large angular size (approximately 100~mas, H03), modest flux density ($\lesssim 1$~Jy, {\it e.g}.\ Frail et al.\ 1996), and close relative angular spacing ($\gtrsim 100~$mas, H03).
The masers also exhibit $\sim 10$\% linear and circular polarization ({\it e.g}.\ C97, H03).
However, several masers may be ``blended'' into a single image feature in observations with angular resolution $\gtrsim 500$~mas, prohibiting the determination of the intrinsic polarization state of the maser emission.
Observations using MERLIN (approximately 100-milliarcsecond [mas] beam) and the VLBA (approximately 10~mas beam) are required in order to determine the intrinsic emission properties of OH (1720~MHz) SNR masers.

In a recent paper (Hoffman et al.\ 2005, hereafter Paper I), we presented full-polarization MERLIN and VLBA observations of approximately 30 OH (1720~MHz) SNR masers in W28.
The observations of W28 in Paper I were shown to be relatively free from confusion due to spatial or spectral blending.
Thus, the intrinsic polarization properties of the maser emission were determined, indicating that:
(1) the magnetic field orientation in the plane of the sky indicated by the linear polarization state of the maser emission is in good agreement with the magnetic field traced by synchrotron emission the SNR limb and
(2) direct observation of the Zeeman splitting between the right- and left-circularly polarized maser lines in OH (1720~MHz) SNR masers is a reliable indicator of the strength of the magnetic field.
However, it should be noted that although observations using MERLIN and the VLBA provide a comprehensive empirical understanding of the maser emission, the interpretation of the results using existing theoretical models is relatively unconstrained ({\it e.g}.\ Paper I; Gray 2003).
In particular, the theoretical derivation of the angle $\theta$ between the line of sight and the magnetic field in the maser gas is not well constrained.

The W44 SNR also contains a large number of masers ($>$25) compared with most OH-maser SNR's, which typically contain only a few masers ({\it e.g}.\ Brogan et al.\ 2000, hereafter BFGT); an observational study of the W44 masers should yield a statistically significant sampling of OH (1720~MHz) SNR maser polarization properties for comparison with the conclusions of Paper I.
W44 is a SNR of diameter approximately 26~pc in the radiative phase of evolution ({\it e.g}.\ Chevalier 1999).
The remnant has an age of about $2 \times 10^4$~yr and a distance of 3.1~kpc determined from observations of the interstellar absorption against the SNR ({\it e.g}.\ Radhakrishnan et al.\ 1972; Caswell et al.\ 1975) and of the dispersion due to the ionized interstellar medium in the direction of the associated pulsar PSR B1853+01 (Wolszczan et al.\ 1991).
The molecular cloud with which the SNR is interacting has been observed in both \ion{H}{1} and molecular lines ({\it e.g}.\ Wootten et al.\ 1977; Sato 1986; Koo \& Heiles 1995; Seta et al.\ 2004).
The molecular cloud/SNR interaction in W44 appears to have a relatively simple geometry; the molecular cloud lies almost directly between the observer and the SNR (Kodaira et al.\ 1977; Wootten 1977).
Seta et al.\ (2004) suggest that the SNR/molecular cloud interface has corrugations $\sim 1$~pc in length so a transverse shock front (suggested to be necessary for the observation of the OH masers) is present at several locations interior to the SNR limb (see their Fig.~11).
Indeed, Figure 1 shows an image of the synchrotron radiation from W44 with the positions of the OH (1720~MHz) maser regions overlaid.
The masers are observed across most of the face of W44 in contrast to the case of W28 (C97) where the masers occur along the limb bordering the molecular cloud at the northeast of the SNR.
Furthermore, the velocities of the masers in W44 lie within the range of approximately 3~\kms\ (a few linewidths) in contrast to the approximately 11~\kms\ ($> 10$ linewidths) over which the W28 masers are observed.
The geometry and velocity structure of the W44 SNR/molecular cloud interaction appears to have been shaped by the creation of a wind-blown cavity prior to the supernova explosion ({\it e.g}.\ Koo \& Heiles 1995).

The OH (1720~MHz) SNR masers in W44 were previously observed by C97 at 1\arcsec\ resolution in an extensive (25 masers), full-polarization experiment using the VLA and by Claussen et al.\ (1999, hereafter C99).
The maser emission was also observed in a comparatively small-scale (2 masers), circular-polarization experiment using MERLIN and the VLBA.
In this paper, we continue the investigation of the polarization of OH (1720~MHz) SNR masers with full-polarization observations of most of the OH masers in W44 using the Multi-element Radio Linked Interferometry Network\footnote{MERLIN is operated as a National Facility by the University of Manchester, Jodrell Bank Observatory, on behalf of Particle Physics and Astronomy Research Council (PPARC).} (MERLIN) and the Very Long Baseline Array (VLBA) of the NRAO\footnote{The National Radio Astronomy Observatory (NRAO) is a facility of the National Science Foundation operated under a cooperative agreement by Associated Universities, Inc.}.
This paper will use the source numbering convention of Claussen et al.\ (C97 Table~1).

\section{Observations}

\subsection{MERLIN+Lovell}

The W44 masers were observed on 26 and 27 January and 28 February 2002 using the MERLIN radio telescope of Jodrell Bank Observatory for a total of $\sim 10$ hours.
The observations used two array tracking positions: (1) the E and F maser region and (2) the B and C maser region (C97); neither the A or D region masers were observed using MERLIN due to time constraints.
The observations were centered on a Doppler velocity of $v_{\rm LSR} = 45.5$~\kms\ for a line rest frequency of 1720.52998~MHz.
Seven antennas were used; the Mark II and Lovell telescopes at Jodrell Bank, the 32~m antenna at Cambridge, and the 25~m dishes at Knockin, Darnhall, Tabley, and Defford.
The very short ($< 500$~m) Mark II-Lovell baseline was not used.
The baseline lengths of MERLIN range from 11~km to 217~km; the array is not sensitive to angular scales larger than 3\farcs3; therefore no continuum radiation from the SNR was detected.
The antennas have right- (R) and left- (L) circularly polarized feeds from which RR, LL, RL, \& LR cross-correlations were formed.
The correlator produced 256 spectral channels across 250~kHz (44~\kms).
The spectra were Hanning-smoothed off-line yielding a velocity resolution of 0.36~\kms.
The visibilities were integrated for 8.0~s.

The absolute amplitude calibration is based on observations of 3C~286.
The bandpasses were calibrated based on observations of 3C~84.
The phases were calibrated using frequent observations (with a bandwidth of 14~MHz) of the source 1904+013 (J1907+0127), which has a flux density of 0.3~Jy and a positional uncertainty of 8~mas.
The uncertainty in the absolute positions of the masers is about 10~mas (depending on the signal-to-noise of the individual detections).
The visibility phases and amplitudes were then (self-) calibrated using images of the E11 maser (approximately 3~Jy, C97).
3C~84 was observed at both 14~MHz and 256~kHz bandwidth allowing bandwidth-dependent calibration to be transferred.
The position angle of the linear polarization response of the antennas was determined from observations of 3C~286 and 1904+013.
Linear-polarization calibration sources were not observed for the B and C region masers.
The RMS noise in a line-free, single channel image is 20~\mjb, in agreement with expected instrumental behavior.
The FWHM synthesized beam of the images is $240\times141$~mas at a position angle $24{\arcdeg}$.

\subsection{VLBA+Y1}

The W44 OH (1720~MHz) masers were observed on 13 and 18 April and 9 and 15 May 2003 with the ten antennas of the VLBA plus one antenna of the VLA for a total of approximately 18~hours.
The antennas have right- (R) and left- (L) circularly polarized feeds from which RR, LL, RL, \& LR cross-correlations were formed.
The correlator produced 128 spectral channels across a 62.5~kHz (11~\kms) band yielding a 0.09~\kms\ velocity channel spacing.
The observing band was centered on $v_{\rm LSR} = 45.425$~\kms.
The spectra were off-line Hanning-smoothed yielding a velocity resolution of 0.18~\kms.
The visibilities were integrated for 7.3~s.

The amplitude scale is set using online system temperature monitoring and {\it a priori} antenna gain measurements.
Bandpass responses and station delays were found from observations of 3C~345.
The phases were calibrated using frequent observations (fast switching) of the source J1838+0404, which is approximately 5\arcdeg\ in angular distance from the masers with a flux density of 0.3~Jy and with a positional uncertainty of 3~mas.
The uncertainty in the absolute positions of the masers is approximately 3~mas (depending on the signal-to-noise of the individual detections).
The visibility phases and amplitudes were then (self-) calibrated using images of the E11 maser (6.6-Jy flux density observed with the VLA by C97; 1.2-Jy flux density observed with the VLBA by C99, see also \S\S3.1, 4.1).
The position angle of the linear polarization response of the antennas is determined from observations of 3C~286 and J1751+0939.

The baseline lengths of the VLBA+Y1 array range from 52~km to 8611~km; the array is not sensitive to angular scales larger than 0\farcs70.
The correlated field of view with respect to the phase-tracking position is $15\arcsec$.
Thus, in order to observe the W44 masers, two array-pointing positions and five correlator-tracking positions were used, summarized in Table~1.
The A region maser was not observed with the VLBA due to time constraints.
Also, the data were significantly contaminated with radio-frequency interference, especially at the Brewster and Saint Croix stations, causing decreased angular resolution in the northwest-southeast direction and decreased image sensitivity.
The severe anisotropy of the angular response of the array was partially remedied through weighting (tapering) of the visibility amplitudes in the $u,v$-plane ({\it e.g}.\ Briggs, Schwab, \& Sramek 1999).
The synthesized beam of the resulting images is $27 \times 13$~mas at a position angle of $-23{\arcdeg}$.
The RMS noise in the final single-channel images is 25~\mjb, in agreement with expected instrumental behavior.

\section{Results}

\subsection{Angular Structure}

All calibration and imaging was performed with the AIPS\footnote{The Astronomical Image Processing System (AIPS) is documented at {\tt http://www.nrao.edu/aips/}.} software package.
Figure \ref{e11_merlin} shows the MERLIN image of the E11 maser region.
Figure \ref{f24_vlba} shows the VLBA image of the F24 maser region.
Tables \ref{mersizes} and \ref{vlbasizes} list the deconvolved angular sizes and total flux densities observed using MERLIN and the VLBA.

All of the five masers observed using MERLIN for which the peak intensity in VLA observations (C97) is greater than 1~\jb\ were detected (B7, E11, E20, F23, and F24).
In addition, the C8 maser (700~\mjb\ with the VLA) was detected in the MERLIN data.
This flux density limited detection rate is consistent with the MERLIN observations of W28 (Paper I).
The C8, E11, and E20 masers have multiple angular components in the MERLIN images ({\it e.g}.\ Fig.~\ref{e11_merlin}), consistent with the observations of the E11 maser by C99 using MERLIN.
Similarly in W28 (Paper I), six masers in the MERLIN data were shown to have multiple components with relative angular separations $\lesssim 1\arcsec$.
The E20 and F23 masers were not detected in the MERLIN observations by C99, as discussed in \S\ref{past}.

The current observations using the VLBA yield detections of two of the ten masers detected using MERLIN, for a detection rate of 20\% consistent with previous observations of OH masers in other SNR's (H03; Paper I).
In addition, the D18 maser, which was not observed with MERLIN, is detected using the VLBA.
The current VLBA observations of the E11A maser are consistent with the VLBA observations of C99.
The current VLBA observations of the F24 maser (Fig.~\ref{f24_vlba}) indicate an extended (approximately 80~mas) angular structure.
The F24 maser was not detected in the VLBA observations by C99, as discussed in \S\ref{past}.

The deconvolved sizes of the MERLIN maser images are in the range 50 to 350~mas (150 to 1000~AU).
Compact cores with deconvolved sizes of 20~mas (60~AU) are apparent in the VLBA images.
These findings are consistent with previous observations of OH SNR masers (C99; Paper I; H03).

\subsection{Line Profiles}

Tables \ref{merpos} and \ref{vlbapos} list the fitted positions and deconvolved spectral-profile properties of the maser lines from the MERLIN and VLBA data.
All line profiles were fitted using the GIPSY\footnote{The Groningen Image Processing System (GIPSY) is described at {\tt http://www.astro.rug.nl/${\mathtt \sim}$gipsy/}.} software package.
Figure \ref{e11_merlin_4pol} shows the full-Stokes spectra for the E11A maser from the MERLIN data.

The line widths determined from the VLBA data are narrower than the line widths determined from the MERLIN data, a trend also observed in the IC~443 and W28 OH masers (H03; Paper I).
(The trend of observed line width decreasing with beam size is observed for many maser species, {\it e.g}.\ OH [4765~MHz]: Palmer, Goss, \& Devine 2003; H$_2$CO [4830~MHz]: Hoffman et al.\ 2003b.)
We assume that for OH (1720~MHz) SNR masers, since the VLBA data appear free from blending (see Paper I), the values for line width derived from VLBA data represent the intrinsic line widths of the masers.
The spectra observed at the F24I and F24II image peaks indicate two velocity components, as summarized in Table~\ref{merpos}.

Figure \ref{e11_merlin_resid} shows the residuals from the best-fit Gaussian to the Stokes $I$ profile of the E11A maser from the MERLIN data.
The line shapes of the W44 OH masers deviate from a Gaussian shape as discussed in \S\ref{PIDisc}.
Also listed in the tables are the brightness temperatures $T_B$ of the masers, determined using the relation $T_B = (0.565 \lambda^2 S)/(k \Theta_a \Theta_b)$, where $S$ is the flux density of the maser (${\rm W}\,{\rm m}^{-2}\,{\rm Hz}^{-1}$), $\lambda$ is the wavelength of the radiation (m), $k$ is Boltzmann's constant, and $\Theta_a$ and $\Theta_b$ are the major and minor axes of the source (radians) ({\it e.g}.\ Rybicki \& Lightman 1979).

\subsection{Zeeman Circular Polarization}

There is sufficient signal-to-noise in the current MERLIN data to measure the magnetic field in the maser gas using two different methods of Zeeman analysis.
The two methods described in this section will be compared in \S\ref{ZeeDisc}.

The Zeeman splitting in the W44 OH masers has been directly resolved as a velocity (frequency) difference between the right-circular polarization (RCP, using RR cross-correlations) and left-circular polarization (LCP, using LL cross-correlations) maser spectra.
The frequency difference is converted to magnetic field using the relationship $v_{\rm RCP} - v_{\rm LCP} = Z B_{\rm RL}$, where $B_{\rm RL}$ denotes the magnetic field strength measured using the frequency separation between the fitted RCP and LCP line profiles and $Z$ is the Zeeman splitting coefficient (see Paper I for a complete discussion of the Zeeman fitting techniques).
As in Paper I, the value $Z = 0.114\ {\rm km}\,{\rm s}^{-1}\,{\rm mG}^{-1}$ ($0.654\ {\rm Hz}\,\mu{\rm G}^{-1}$; Davies 1974) is used.
The observed velocity separation of the circularly-polarized maser components may be parametrized with $x_B \equiv \Delta{\nu_B}/\Delta{\nu_{\rm D}}$, where $\Delta{\nu_B}$ is the Zeeman frequency separation of the split substates and $\Delta{\nu_{\rm D}}$ is the Doppler-broadened width of the maser line. 
For the W44 maser components $x_B \approx 0.1$ (Table \ref{circ}).

The following equation relates the magnetic field to the Stokes $I$ and Stokes $V$ line profiles ({\it e.g}.\ Roberts et al.\ 1993; Sarma et al.\ 2001; BFGT),
\begin{equation}
V = b \frac{\partial{I}}{\partial{\nu}} + aI\ .
\label{zeemap}
\end{equation}
The Stokes $I$ and $V$ profiles of the maser lines from both the MERLIN and VLBA observations were fitted to equation~\ref{zeemap}.
All fitting was performed with the MIRIAD\footnote{The Multichannel Image Reconstruction, Image Analysis and Display (MIRIAD) is documented at {\tt http://bima.astro.umd.edu/miriad/}.} software package.
The $aI$ term in equation~\ref{zeemap} represents the symmetrical component of the Stokes $V$ profile.
The fitted magnitude of $a$ (approximately 0.02) is consistent with instrumental ``leakage'' of Stokes $I$ into Stokes $V$; in the current data $a$ does not represent an intrinsic property of the maser radiation.
The values of the fitted parameter $b$ (approximately 0.03) are related to the magnetic field $B_{\theta}$ using $b = Z B_{\theta}/2$ (see also the discussion Paper I and \S4.3 concerning conversions of $b$ to $B$ which involve $\theta$).

Table~\ref{circ} lists the values of $v_{\rm RCP} - v_{\rm LCP}$, $B_{\rm RL}$, $x_B$, and $B_{\theta}$ for the MERLIN observations.
Although the W44 masers are detected in Stokes $V$ using the VLBA, there is insufficient signal to noise ($< 3\sigma$) to fit a magnetic field to the Stokes $V$ profiles.
The sign and magnitude of the magnetic fields fitted to the MERLIN data are consistent with VLA observations (C97).
Figure \ref{bcomp} shows the good agreement between the $B_{\rm RL}$ and $B_{\theta}$ measurements of the magnetic field.

\subsection{Linear Polarization}

Table \ref{merpol} lists the properties of the linear polarization of the masers observed using MERLIN.
The linear-polarization measurements have not been corrected for Faraday effects (see \S4.4.1).
Linear polarization is not detected using the VLBA with a 3-$\sigma$ upper limit of 10\%, consistent with the MERLIN observations which show that the typical linear polarization fraction is approximately 5\% to 10\%.
The linear polarization angles $\chi$ listed in Table \ref{merpol} are determined using the usual relation $\tan{2\chi} = U/Q$, where $U$ and $Q$ are the Stokes parameters formed from the cross correlations of the right- and left-circularly polarized emission.
Figure \ref{e11_merlin_4pol} shows the full-polarization profiles for the E11A maser observed with MERLIN.
The dotted lines in Figure \ref{e11_merlin_4pol} show the fitted profiles using (1) the Stokes $I$ line profile summarized in Table \ref{merpos}, (2) the Stokes $V$ profile fitted using equation \ref{zeemap}, and (3) the fitted Stokes $Q$ and $U$ profiles (Table \ref{merpol}).
The percentage of linearly polarized intensity, $q \equiv \sqrt{Q^2+U^2}/I$, is also listed in Table \ref{merpol}.
The values of $\chi$ and $q$ are in good agreement with those observed by C97 using the VLA.

\section{Discussion}

\subsection{Comparison with Previous Observations of W44\label{past}}

The E and F regions of OH (1720~MHz) SNR maser emission in W44 were previously observed by C99 using both MERLIN+Lovell and the VLBA+Y1.
The beam size ($290 \times 165$~mas at a position angle of 25\arcdeg) and image noise (30~\mjb) of the C99 MERLIN images are comparable to (through slightly larger than) those of the current MERLIN observations ($240 \times 141$~mas at a position angle $24{\arcdeg}$, 20~\mjb).
The VLBA observations discussed by C99, however, were much shorter in duration than the current observations (approximately 3.5~hr on-source compared with approximately 24~hr) and the image noise was correspondingly higher (100~\mjb compared with 25~\mjb).
The C99 VLBA images had a larger beam of $40 \times 30$~mas at a position angle of $-10\arcdeg$ compared with the beam of the current VLBA observations: $27 \times 13$~mas at a position angle of $-23{\arcdeg}$.

Using MERLIN, C99 imaged the E11 and F24 masers in W44.
The angular structure observed in the E11 maser is similar to that shown in the current data (Fig.~\ref{e11_merlin}).
The deconvolved angular size and flux density of the E11A maser are also in agreement between the two observations.
The peak flux densities measured for the F24 maser are in agreement to approximately 30\% (C99: 1.7~\jb; current data: 2.3~\jb) between the two observations.
Two additional masers, E20 and F23, have been detected in the current MERLIN data.
Although these two masers are each about 60\% weaker than the E11 and F24 masers, the E20 and F23 masers, at their current flux densities, should have been detectable in the C99 data.

Using the VLBA, C99 were only able to detect the E11A maser.
In the C99 images the E11A maser is unresolved with a flux density of 1.2~Jy, consistent with the current observations of a slightly resolved source having flux density 1.1~Jy (Table \ref{mersizes}).
Based on the new VLBA observations, there is also a detection of the F24 maser (Fig.~\ref{f24_vlba}).
The F24 maser is not detected in the VLBA data by C99, presumably due to insufficient $u$-$v$ coverage in the C99 observations: the 2~Jy of total flux density in the data of the F24 maser is distributed over three components which subtend greater than 60~mas.
The C99 experiment used an observing schedule which divided eight hours of VLBA observing time between the W44 and W28 masers.
The current VLBA experiment dedicated seven-hour tracks to observations of the W44 masers, resulting in denser $u$-$v$ coverage than the C99 schedule.
No polarization is discussed by C99 for the observations of the W44 masers.

\subsection{Maser Amplification}

The amplification process of an unsaturated maser may be parametrized by the gain $\tau$ such that $T_B \approx T_{bg} e^{-\tau}$, where $T_B$ is the brightness temperature of the maser and $T_{bg}$ is the brightness temperature of the background radiation which provides the ``seed'' radiation for the stimulated emission process.
In this section we discuss constraints on the agreement between the observed brightness temperature of the masers and the maser gain suggested using the collisional pump model.

Elitzur (1976) suggests that the maximum gain of a collisionally pumped 1720-MHz OH maser is $\tau \simeq -20$.
Pavlakis \& Kylafis (1996) suggest that the maximum brightness temperature of a collisionally pumped 1720-MHz OH maser $T_B \simeq 10^9$~K, corresponding to a gain $\tau \simeq -20$.
These suggestions for the value of the gain $\tau \gtrsim -20$ may be compared with measurements of $T_B$ and $T_{bg}$ using $T_B \approx T_{bg} e^{-\tau}$.
$T_{bg}$ is known because of the geometry of the SNR/molecular cloud interaction in W44: the molecular cloud is directly between the observer and the SNR ({\it e.g}.\ Kodaira et al.\ 1977) such that the masers likely amplify the 1720-MHz synchrotron continuum radiation from the SNR.

The brightness temperature of the background radiation at 1720~MHz may be extrapolated from observations at 1442~MHz by Giacani et al.\ (1997, with resolution of 15\arcsec).
For synchrotron radiation $T_B \propto \lambda^{2-\alpha}$ ({\it e.g}.\ Rybicki \& Lightman 1979), where $\alpha$ is the spectral index of the synchrotron radiation, measured by Giacani et al.\ (1997) to be $\alpha \approx -0.4$.
The brightness temperature of the 1442-MHz synchrotron radiation in the E and F maser region is about 90~K, corresponding to a brightness temperature for the 1720~MHz background radiation of $T_{bg} \approx 60$~K.
Therefore, the gain corresponding to a maser brightness temperature $T_B \approx 10^9$~K (Table \ref{vlbapos}) is $\tau \simeq -17$.
If the maser is saturated, this analysis yields the lower limit on the strength of the gain $\tau \lesssim -17$.

For the case in which the masers amplify their own spontaneous emission, rather than an external background source, $T_{bg}$ is determined from the excitation temperature $T_x$ of the spectral line (see, {\it e.g}., Elitzur 1992).
If $|T_x| \gtrsim |T_{bg}|$ the seed provided internal to the masers is greater than the seed provided by the external background, in which case the gain will not be as strong as calculated above, indicating the upper limit $\tau \gtrsim -17$.
In the absence of firm constraints on $T_x$, we assume the approximate value $\tau \simeq -17$.
Since the estimated gain does not exceed the suggested maximum gain, we find these observations to be consistent with suggestions based on the collisional pump model (see also \S4.5.1).

\subsection{Zeeman Detections\label{ZeeDisc}}

The magnetic field in the masers has been measured using both resolved Zeeman splitting between the RCP and LCP line profiles ($B_{\rm RL}$) and fitting of Stokes $V$ to the derivative of Stokes $I$ ($B_{\theta}$).
The values of $B_{\theta}$ are consistent with the values of $B_{\rm RL}$ (Fig.~\ref{bcomp}).
However, despite the good agreement between the derived parameters $B_{\theta}$ and $B_{\rm RL}$, it is important to note that the theoretical models used to interpret the observations may not be appropriate.
Since theoretical investigations of maser polarization are usually based on $x_B > 1$ or $x_B \ll 1$ (cf.\ Elitzur 1996), many of the theoretical models (concerning both Zeeman analysis and linear polarization analysis) may not be applicable to these OH SNR masers for which $x_B \approx 0.1$ (\S3.3; Table \ref{circ}; see also Paper I).

The sign and magnitude of the magnetic fields indicated by the current circular-polarization observations are consistent with previous observations of OH (1720~MHz) SNR masers.
The sign of the fitted magnetic field indicates the direction of the magnetic field projected onto the line of sight (although the magnitude of the Zeeman splitting may not be {\em linearly} related to the line-of-sight component of the field, see Paper I).
All of the W44 masers are fitted with a negative magnetic field indicating a magnetic field oriented toward the observer.
The magnitude of the magnetic field measured using the circular-polarization data, although uncertain by a factor of approximately three due to uncertainties in $\theta$ and the degree of saturation ({\it e.g}.\ Watson \& Wyld 2001, hereafter WW01), may be compared with theoretical suggestions about the strength of the magnetic field in shock-compressed molecular clouds.
As discussed by BFGT, Bertoldi \& McKee (1992) suggest $|B| = 0.4 \Delta{v} n^{0.5}$, where $n$ is the density in ${\rm cm}^{-3}$ and $\Delta{v}$ is the shocked line width in \kms.
Frail \& Mitchell (1998), using observations of CO ($J = 3 \rightarrow 2$), estimate the density in the post-shock to be $n \approx 10^5\ {\rm cm}^{-3}$ and the shocked linewidth to be $\Delta{v} = 40$~\kms.
For these values, the Bertoldi \& McKee (1992) model yields $|B| \approx 5$~mG which is of the same order as the magnetic field magnitude estimated from the observed Zeeman splitting ($\approx 1$~mG).

\subsection{Linear Polarization}

In this section we compare the magnetic field direction indicated by the shock front and the magnetic field direction indicated by the linearly-polarized maser emission.
Using this comparison, we discuss constraints on the magnetic field geometry within the masers.

\subsubsection{Other Observations of Magnetic Field Orientation\label{synch}}

The magnetic field orientation in the SNR may be probed directly using observations of the synchrotron continuum radiation.
Milne \& Dickel (1975) have observed the linear polarization of the synchrotron emission from W44 at both 5~GHz and 2.7~GHz at an angular resolution of 4\farcm4 (see also Milne [1987] which contains a review of the observations by Kundu \& Velusamy [1972] and Dickel \& Milne [1976]).
Using a dual-frequency analysis, they find the Faraday rotation measure (RM) in the direction of the masers to be between 0 and $-10$ ${\rm rad}\,{\rm m^{-2}}$ and the depolarization to be negligibly small.
Based on these synchrotron observations, we conclude that the Faraday rotation and depolarization affecting our maser measurements is small and need not be considered.
Milne and Dickel also find the position angle of the plane-of-the-sky component of the magnetic field at the location of the masers to be approximately $-35\arcdeg$.
The magnetic field orientation is uniform in direction for several arcminutes on every side of the maser region.
Based on the relative uniformity of the results of the arcminute-resolution synchrotron observations, we assume that the Faraday rotation, depolarization, and magnetic field orientation do not change appreciably over the $\lesssim 1\arcmin$ angular separation of the masers.

The coincidence of OH SNR masers with shocked molecular gas is well established ({\it e.g}.\ Frail \& Mitchell 1998).
As the shock forms and moves outward through the interstellar medium, the ambient galactic magnetic field is swept up and compressed.
For example, Balsara et al.\ (2001) suggest that the direction of the magnetic field in the shocked gas lies in a plane parallel to the shock.
There is a good agreement between the orientation of the magnetic field inferred using images of the shocked CO gas (Frail \& Mitchell 1998) and the position angle of the magnetic field measured from observations of the synchrotron radiation ($-35\arcdeg$, Milne \& Dickel 1975).

\subsubsection{Orientation of the Linear Polarization of the Masers}

The position angle $\chi$ of the linearly-polarized emission from the OH masers in W44 (Table \ref{merpol}) is parallel to the position angle of the magnetic field apparent from both the CO and synchrotron observations.
The agreement between the orientation of the CO shock front and the maser position angle is shown in Figure \ref{CO}.
From these comparisons, it is apparent that the position angle of the linear polarization of the maser is parallel to the plane-of-the-sky component of the magnetic field in the maser gas.

Based on theoretical investigations, Elitzur (1998) and Watson and Wyld (WW01), for example, suggest that the polarization fraction $q$ and the polarization angle $\chi$ of maser radiation depend strongly on $\theta$, the angle between the magnetic field and the line of sight.
For masers viewed at an angle of $\theta \lesssim 55\arcdeg$, these authors suggest that the linear polarization angle is parallel to the plane-of-the-sky component of the magnetic field permeating the maser gas.
For masers viewed at an angle of $\theta \gtrsim 55\arcdeg$, the linear polarization angle is suggested to be perpendicular to the plane-of-the-sky component of the magnetic field.
The OH masers in W44 appear to have a linear polarization parallel to the plane-of-the-sky component of the magnetic field in the SNR shock, implying $\theta \lesssim 55\arcdeg$.
In contrast, the OH masers in W28 appear to have a linear polarization perpendicular to the magnetic field of the SNR shock (Paper I).

The masers are observed to have $q \approx 10$\% (Table \ref{merpol}; see also Paper I; BFGT).
For the case in which $q \approx 10$\% and in which $\chi$ is parallel to the plane-of-the-sky component of the magnetic field, Watson and Wyld (WW01), for example, suggest that $\theta \approx 55\arcdeg$.
Indeed, all observations of the linear polarization of OH (1720~MHz) SNR masers interpreted using the theories of WW01 and Elitzur (1998) yield $\theta \approx 55\arcdeg$ (see BFGT; Paper I).

Thus, the W44 masers and W28 masers differ in the relative orientation of the maser polarization and the magnetic field, for which theoretical polarization models suggest $\theta \gtrsim 55\arcdeg$ in W28 and $\theta \lesssim 55\arcdeg$ in W44.
However, since the fraction of linear polarized maser emission $q$ is comparable in both the W44 and W28 OH SNR masers, $\theta$ must be approximately $55\arcdeg$ in both remnants.
We consider the preponderance of the derived value $\theta \approx 55\arcdeg$ to represent an uncomfortable level of coincidence in the theoretical interpretations.
We plan to discuss this coincidence further in a future paper.
Also, observations with more favorable spectral resolution will more accurately determine line shape parameters, such as the deviation from Gaussian (Watson et al.\ 2002) and the relative width of Stokes $I$ and $V$ ({\it e.g}.\ Vlemmings et al.\ 2002), which are suggested to depend upon the degree of saturation of the masers and $\theta$, the angle between the line of sight and the magnetic field.

\subsection{Comparison with Discussion in Paper I\label{PIDisc}}

\subsubsection{Maser Gain Saturation}

Three separate tests indicate partial saturation of the masers.
First, theoretical investigations indicate that a maser must be at least partially saturated in order to possess linear polarization, assuming no anisotropies
in the pumping ({\it e.g}.\ Elitzur 1998; WW01).
The degree of linear polarization observed in the W28 and W44 masers ($q \approx 10$\%) is consistent with partial saturation (Paper I).
Second, the estimated gain of the masers, $\tau \simeq -17$ (\S4.2), is comparable to the maximum suggested gain for OH SNR masers, consistent with the conclusion that the masers are partially saturated.
Third, Watson et al.\ (2002) suggest the line-shape parameters $\delta$ (defined therein) and kurtosis ({\it e.g}.\ Abramowitz \& Stegun 1972; Press et al.\ 2001) are dependent upon the saturation state of the maser amplification process.
The deviations from Gaussian line shape observed in the W44 maser profiles ({\it e.g}.\ Fig.~\ref{e11_merlin_resid}) are similar to those observed for the W28 masers ($\delta \sim 10^{-3}$ and a positive kurtosis; note that Paper I inaccurately described the kurtosis as negative).
This result is consistent with partial saturation (though it is also consistent with unsaturated masing).
We intend to make a more quantitative analysis using the Watson et al.\ (2002) model in a future paper.

\subsubsection{Zeeman Analysis}

In Paper I we presented the first observation of Zeeman splitting in OH (1720~MHz) SNR masers to be resolved directly between the RR and LL line profiles.
Both the MERLIN and VLBA observations of the W28 masers yield a resolved velocity difference between Zeeman components.
Due to the relatively low flux density of the W44 masers, only the MERLIN data in this paper indicate a difference between the RR and LL line centers (Table \ref{circ}).
Nonetheless, for both the W44 and W28 observations, the Zeeman effect has been directly resolved between the RR and LL polarizations.

A comparison of the magnetic field strength calculated from the RR and LL line separation ($B_{\rm RL}$) and that calculated from the Stokes-fitting method ($B_{\theta}$) indicates, for both the W44 and W28 masers, that $B_{\rm RL}/B_{\theta} \approx 1$ (Fig.~\ref{bcomp}).
For both the W28 and W44 masers, the ratio of line splitting to line width is $x_B \approx 0.1$.
Since $B_{\theta}$ is expected to depend upon $\theta$ but $B_{\rm RL}$ is not, we conclude that their agreement indicates that either (1) the magnitude of the magnetic field fitted to the polarization observations does not depend as strongly on $\theta$ as is currently suggested in the theoretical literature ({\it e.g}.\ WW01) or (2) the observed Zeeman splitting in the masers ($x_B \approx 0.1$) renders models based upon $x_B \ll 1$ inapplicable to these results.
We plan to collect all polarization observations of OH SNR masers into a future paper which addresses the appropriateness of the theoretical interpretations.

\section{Conclusions}

Using MERLIN, we have detected all but one of the OH (1720~MHz) SNR masers in W28 with intensity greater than 1~\jb\ as observed with the VLA by Claussen et al.\ (1997).
Of the 10 MERLIN detections, we have detected two masers (approximately 20\%) with the VLBA+Y1, consistent with other VLBI detection rates of OH (1720~MHz) SNR masers (H03; Paper I).
Based on MERLIN data having a resolution of 200~mas and VLBA data with a resolution of 15~mas, the masers are measured to have deconvolved angular sizes of 50 to 350~mas (150 to 1000~AU) with compact cores 20~mas (60~AU) in size, consistent with theoretical expectation and previous observations.
All of the imaging results for the OH (1720~MHz) SNR masers in W44 are consistent with theoretical expectation and previous observations.
In general, OH (1720~MHz) SNR masers which are not affected appreciably by interstellar scattering (see Yusef-Zadeh et al.\ 1999a; Claussen et al.\ 2002) have similar sizes and angular morphologies (Paper I; H03).

Zeeman splitting is observed between the right- and left-circularly polarized maser lines, offering a direct measurement of the magnetic field in the maser gas.
The strengths of the magnetic fields estimated from the circular-polarization observations ($\sim 1$~mG) are consistent with the magnetic field strengths suggested using models of shock-compressed molecular clouds.
Also, the position angle of the magnetic field measured using the linear polarization of the masers is in good agreement with both the orientation of the shock observed using molecular emission ({\it e.g}.\ Frail \& Mitchell 1998), with the magnetic field orientation expected from the SNR expansion ({\it e.g}.\ Balsara et al.\ 2001), and with the magnetic field orientation calculated from observations of the synchrotron emission ({\it e.g}.\ Dickel \& Milne 1975).
Thus, the linear polarization angle of the W44 OH masers appears to be parallel to the SNR magnetic field in contrast to the position angle of the linear polarization of the W28 OH masers (Paper I) which was found to be perpendicular to the magnetic field in the SNR.
In addition, interpretation using existing theoretical models (appropriate for $x_B \ll 1$ although the masers have $x_B \approx 0.1$) indicate that $\theta \approx 55\arcdeg$, as has been found for all other OH-SNR-maser polarization observations.

\acknowledgments

IMH is supported by the NRAO pre-doctoral researcher program.
We thank W.\ D.\ Watson and M.\ Elitzur for helpful discussions concerning the interpretation of the maser theories.
We thank A.\ M.\ S.\ Richards for assistance with reduction of the MERLIN observations and P.\ Thomasson for flexible scheduling of the MERLIN array.

\clearpage

\begin{figure}
\plotone{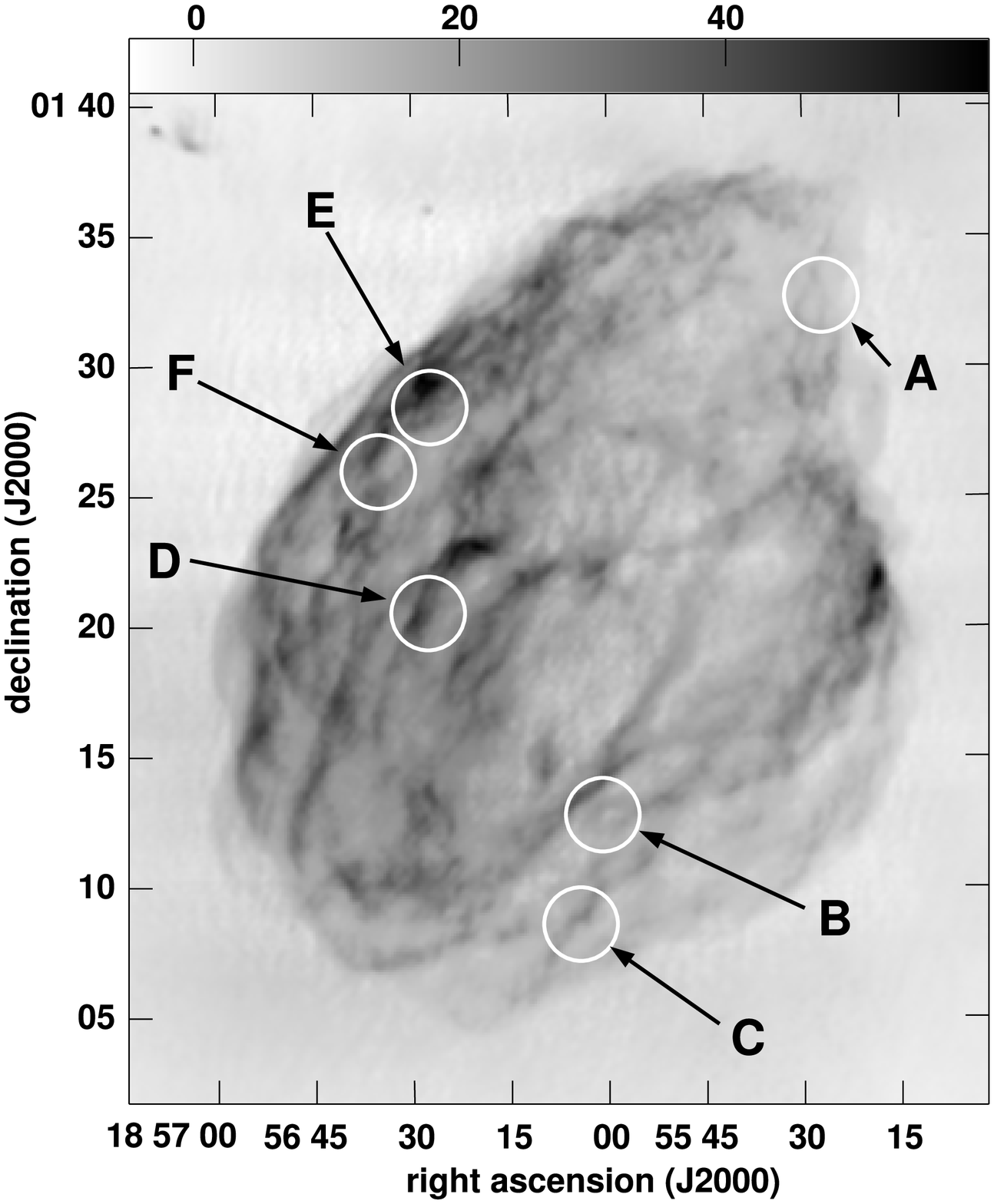}
\caption{A greyscale image of the 1442-MHz synchrotron continuum radiation from W44 (Giacani et al.\ 1997).
The scale at the top of the figure is in units of \mjb.
The image has a resolution of $16\arcsec \times 15\arcsec$.
The location of the various regions of OH (1720~MHz) emission concentrations are indicated (after Claussen et al.\ 1997).
The sizes of the ellipses used to to indicate the maser regions have no special significance; the regions contain between one and ten masers as observed with the VLA (C97).
Using MERLIN we observed the masers in regions B, C, E, and F; using the VLBA we observed the masers in regions B, C, D, E, and F.
\label{fig1}}
\end{figure}

\clearpage

\begin{figure}
\plotone{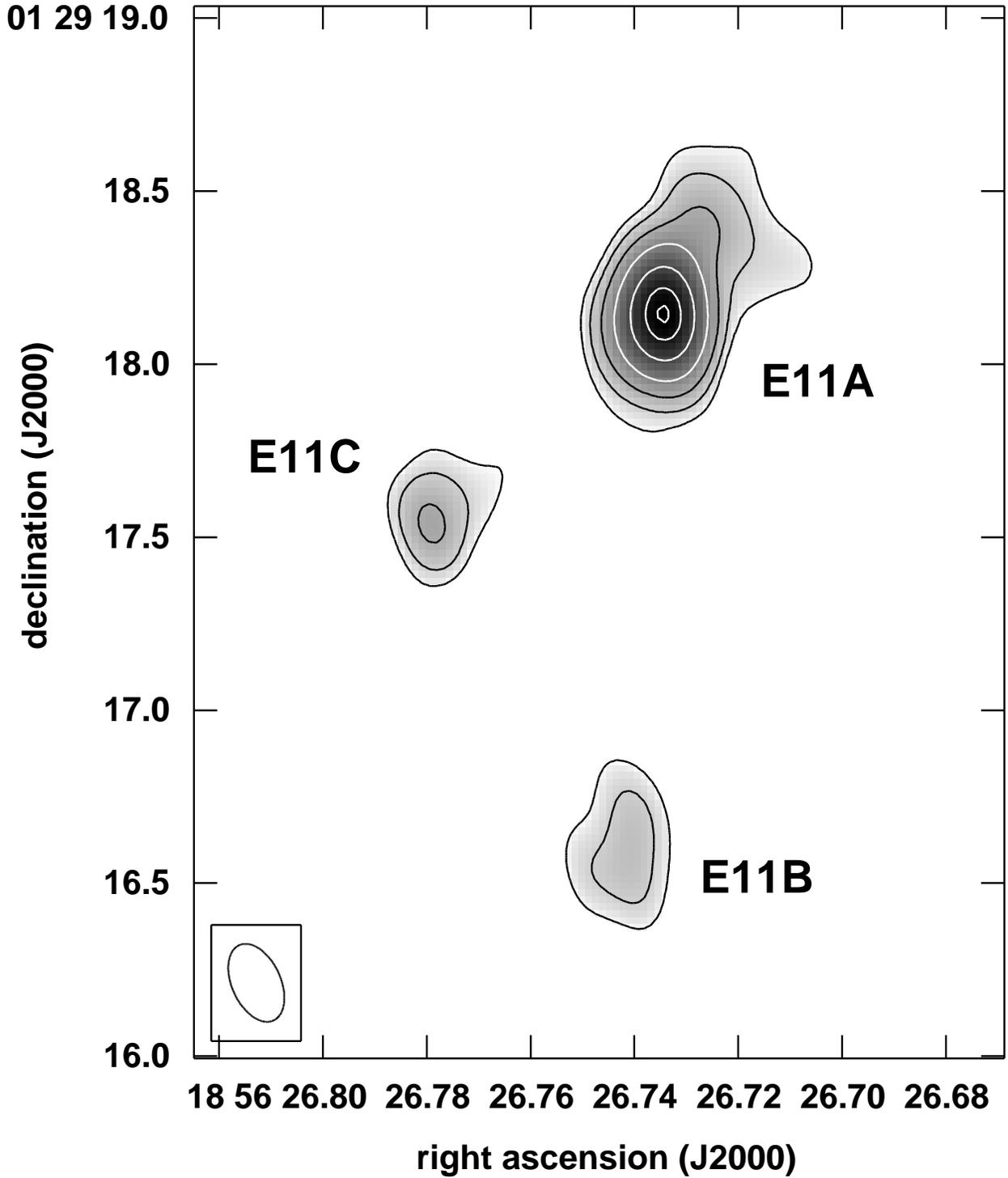}
\caption{Contour and greyscale image of the E11 maser region at $v_{\rm LSR} = 44.2$~\kms\ from the MERLIN observations.
The contour levels are -4, 4, 8, 16, 32, 64, 100, and 120 times the image RMS noise level of 20~\mjb\ (no negative contours appear).
The beam, plotted in the lower left-hand corner, is $240 \times 141$~mas at a position angle of 24\arcdeg.
\label{e11_merlin}}
\end{figure}

\clearpage

\begin{figure}
\plotone{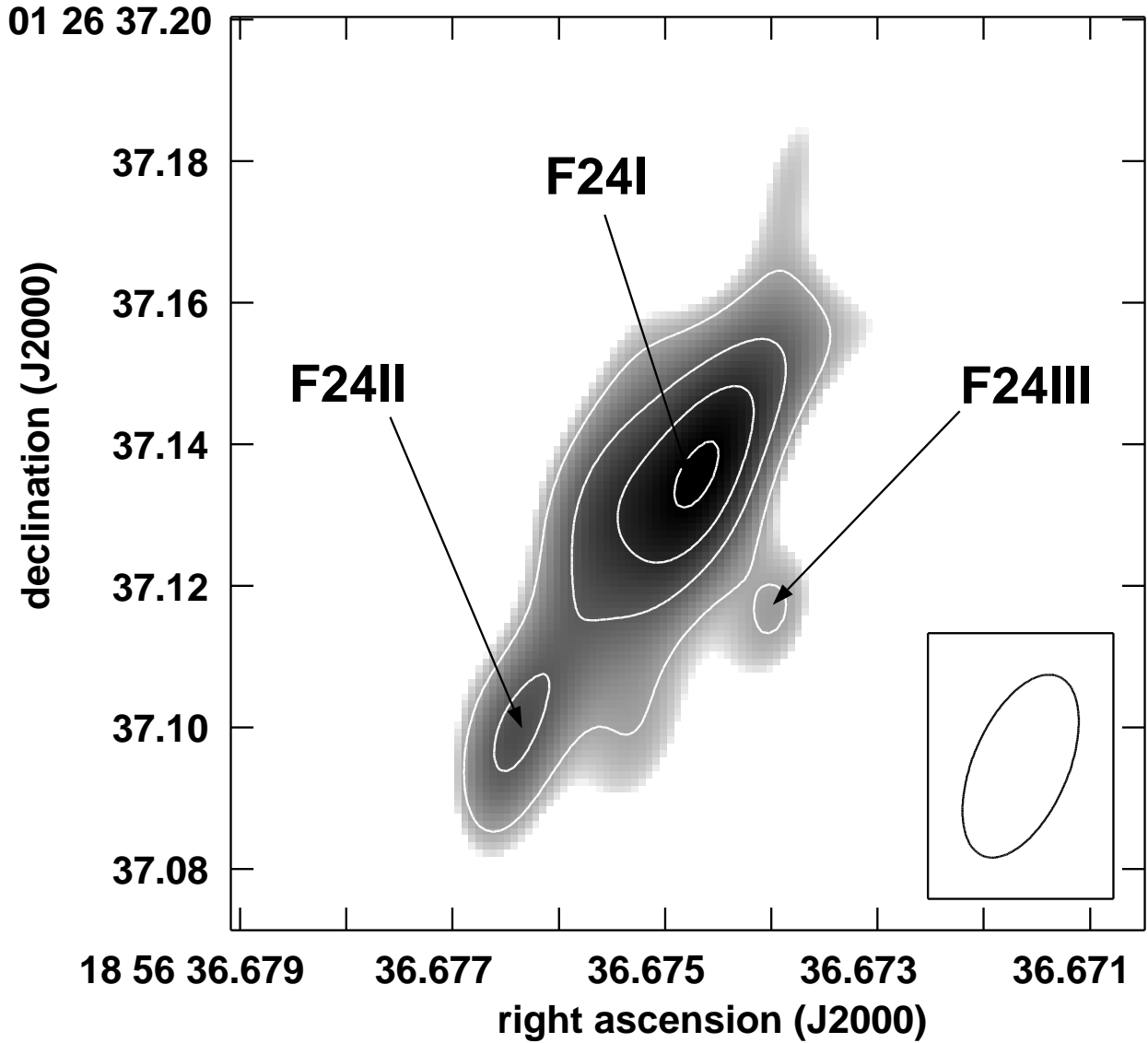}
\caption{Contour and greyscale image of the F24 maser region at $v_{\rm LSR} = 46.8$~\kms\ from the VLBA observations.
The contour levels are -7, 7, 14, 21, and 28 times the image RMS noise level of 25~\mjb\ (no negative contours appear).
The beam, plotted in the lower right-hand corner, is $27 \times 13$~mas at a position angle of $-23$\arcdeg.
\label{f24_vlba}}
\end{figure}

\clearpage

\begin{figure}
\plotone{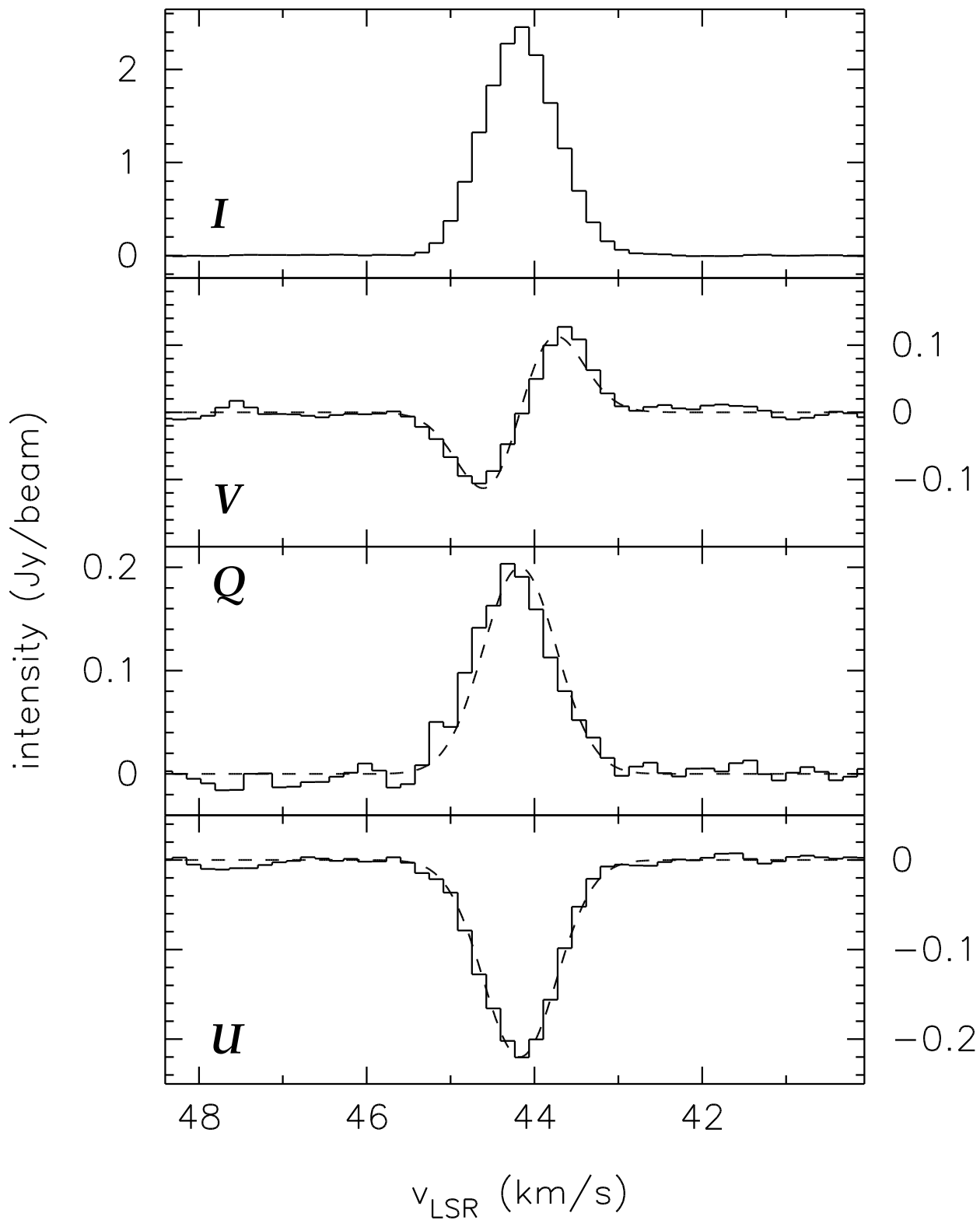}
\caption{Full stokes spectra from the MERLIN observations for the E11A maser.
The dashed line in the Stokes $V$ panel is the Stokes $I$ derivative scaled by a $-0.61$-mG magnetic field (Table \ref{circ}).
The dashed lines in the Stokes $Q$ and $U$ panels are the best-fit Gaussians to the line profiles (Table \ref{merpol}).
\label{e11_merlin_4pol}}
\end{figure}

\clearpage

\begin{figure}
\plotone{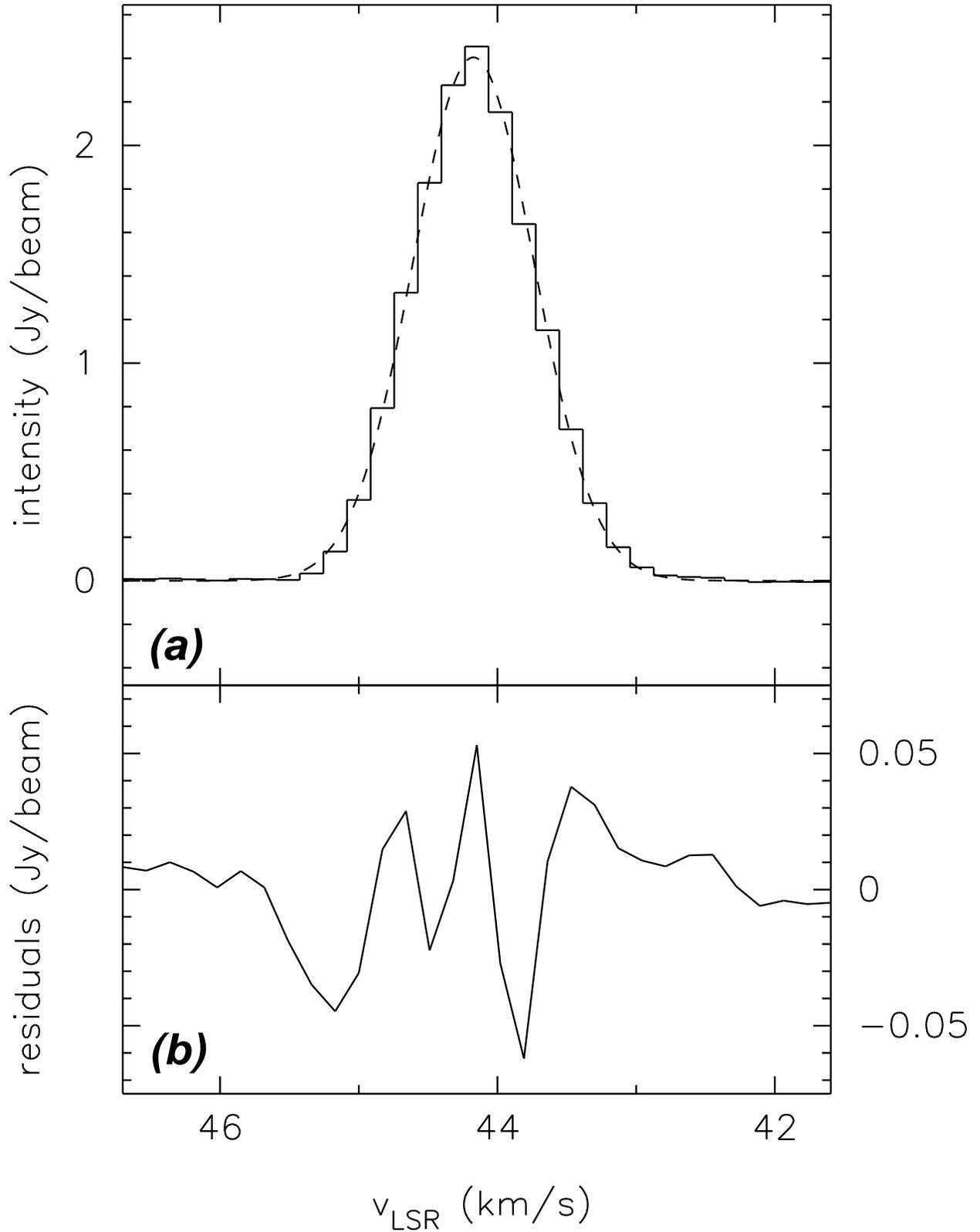}
\caption{{\it (a) (solid line)} The Stokes $I$ line profile of the E11A maser from the MERLIN observations.
{\it (dashed line)} The best-fit Gaussian to the line profile (Table \ref{merpos}).
{\it (b)} The residuals from the best-fit Gaussian, discussed in \S4.5.1.
\label{e11_merlin_resid}}
\end{figure}

\clearpage

\begin{figure}
\plotone{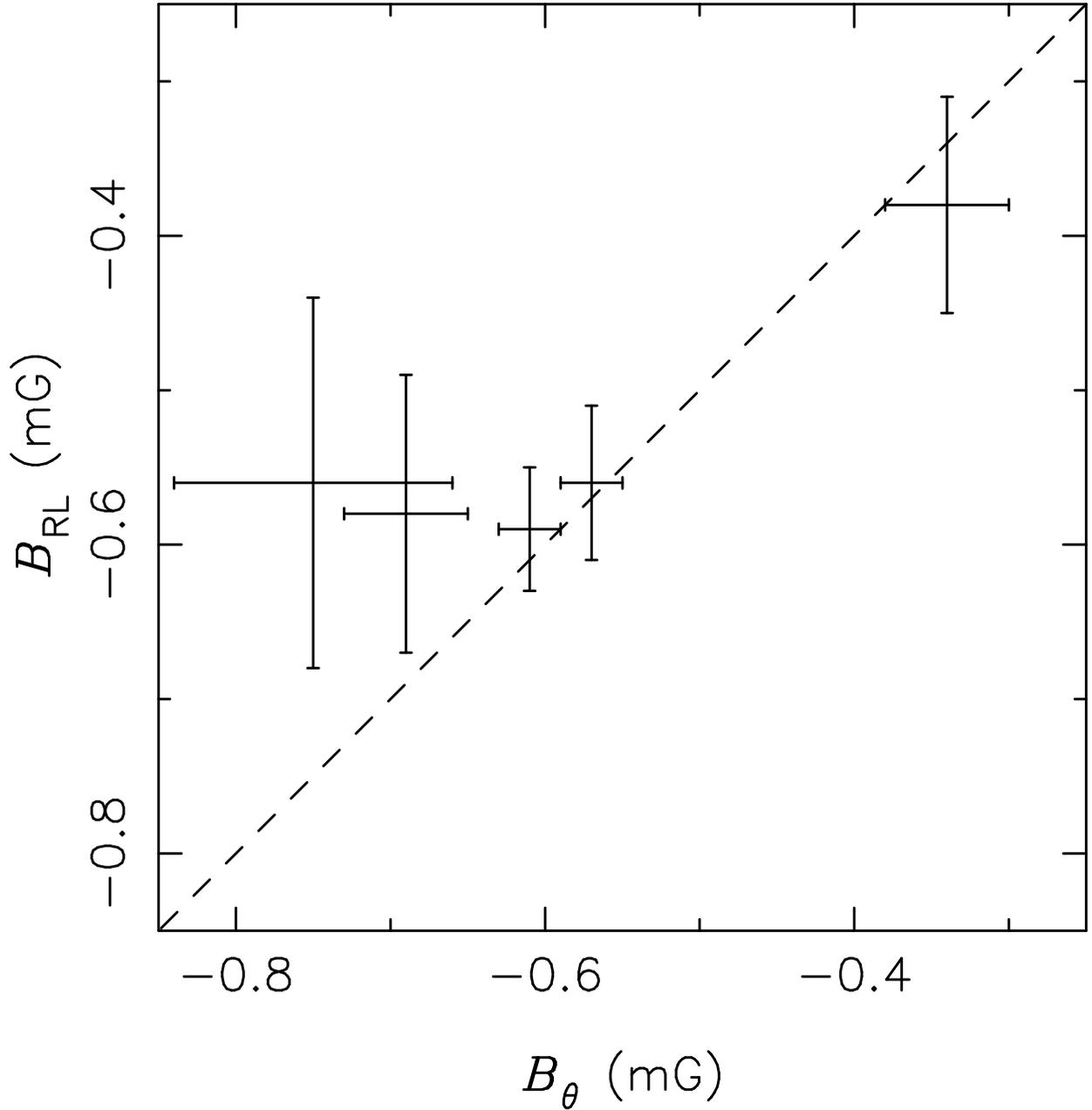}
\caption{A plot of the magnetic field strengths obtained for the masers in the MERLIN observations (Table \ref{circ}) using $V/(\partial{I}/\partial{\nu})$-fitting ($B_{\theta}$) and RCP-minus-LCP fitting ($B_{\rm RL}$), as described in \S3.3.
The dashed line has a slope of one indicating the good agreement between the two measurement methods.
\label{bcomp}}
\end{figure}

\clearpage

\begin{figure}
\plotone{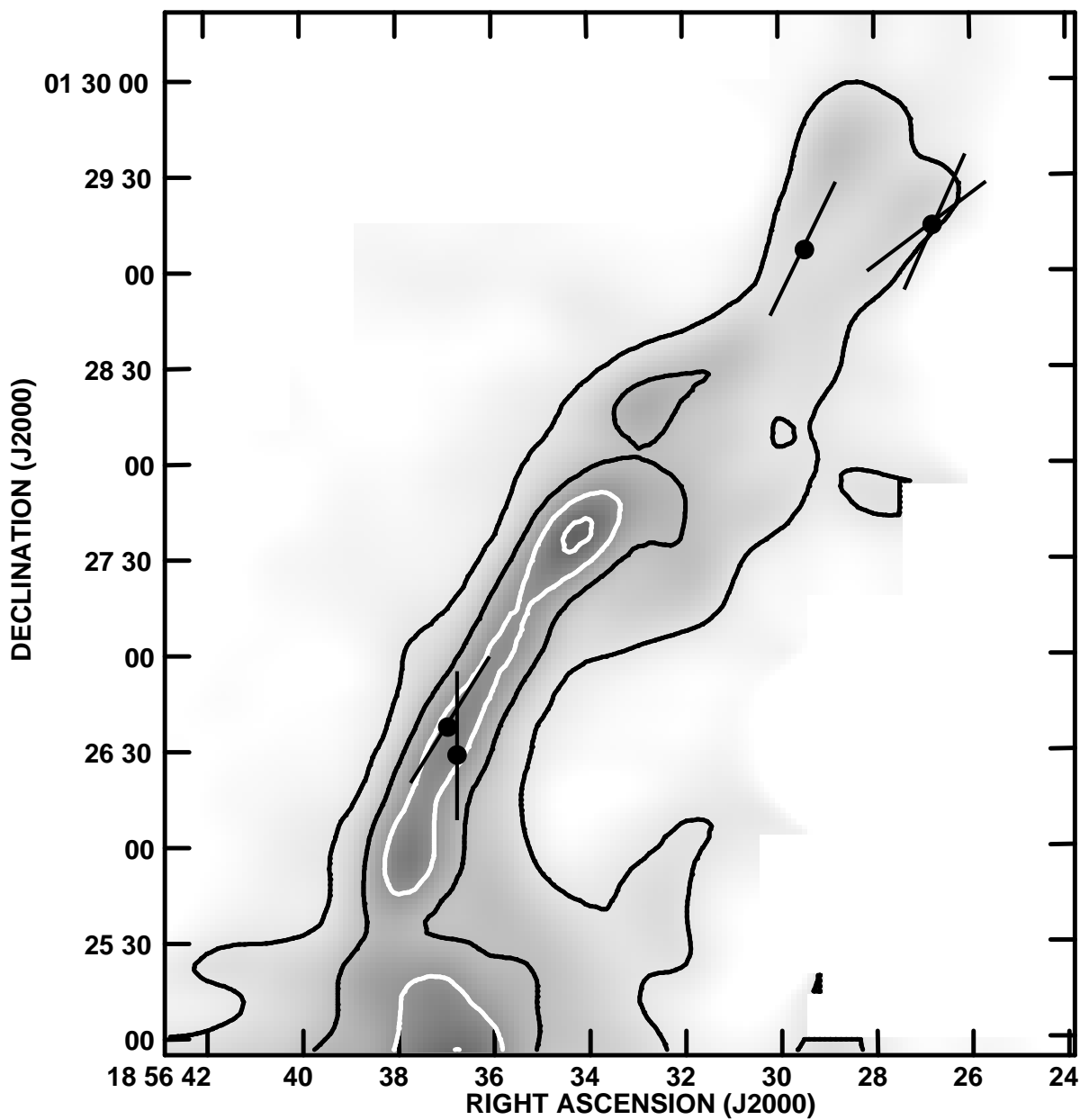}
\caption{Contour and greyscale image of the CO ($J=3 \rightarrow 2$, 346~GHz) line emission at $v_{\rm LSR} = 45$~\kms\ from the W44 E and F maser regions observed using the James Clerk Maxwell Telescope (C.\ L.\ Brogan 2004, private communication).
The contours are at 3, 5, 7, and 9 times the image RMS noise level of 9~K.
The beam of the telescope is 15\arcsec.
The filled circles mark the positions of the OH masers for which the position angle of the linear polarization has been determined using MERLIN (Table \ref{merpol}); the straight lines through the maser markers indicate the linear polarization angle.
The good agreement between the orientations of the shocked CO gas and the maser polarization is discussed in \S \ref{synch}.
\label{CO}}
\end{figure}

\clearpage

\begin{deluxetable}{ r@{\ }r@{\ }l  r@{\ }r@{\ }l }
\tablecaption{VLBA Observation Positions}
\tablecolumns{6}
\tablewidth{0pt}
\tablehead{
\multicolumn{3}{c}{R.A.\ (J2000)} & \multicolumn{3}{c}{Decl.\ (J2000)} \\
\colhead{(h} & \colhead{m} & \colhead{s)} & \colhead{(\arcdeg} & \colhead{\arcmin} & \colhead{\arcsec)}
}
\startdata
\cutinhead{array tracking positions}
18&56&01.4902 &  01&10&42.177 \\
  &  &32.2215 &    &24&59.361 \\
\cutinhead{correlation centers}
18&56&01.2889 &  01&12&46.784 \\
  &  &03.8015 &    &08&46.949 \\
  &  &28.1418 &    &29&12.076 \\
  &  &29.4822 &    &20&26.555 \\
  &  &36.6571 &    &26&33.176 \\
\enddata
\end{deluxetable}

\clearpage

\begin{deluxetable}{ l  r@{.}l  r@{.}l@{$\times$}r@{.}l c }
\tablecaption{MERLIN Maser Sizes and Flux Densities\label{mersizes}}
\tablecolumns{8}
\tablewidth{0pt}
\tablehead{
\colhead{Feature} & \multicolumn{2}{c}{$S$ (Jy)} & \multicolumn{4}{c}{angular size\tablenotemark{a}\ (mas)} & \colhead{P.A.\ (deg)}
}
\startdata
B7     &  0&9  & 0&24(2)   &$<0$&07    &173(4)   \\
C8A    &  0&4  & 0&23(5)   &$<0$&17    &148(21)  \\
C8B    &  0&3  & 0&34(7)   &$<0$&14    &171(9)   \\
E11A   &  2&7  & 0&20(1)   &$<0$&05    &150(3)   \\
E11B   &  0&2  & 0&37(6)   &  0&15(5)  &166(12)  \\
E11C   &  0&4  & $<0$&19   &$<0$&15    &144(18)  \\
E20A   &  1&5  & 0&35(1)   &$<0$&07    &155(2)   \\
E20B   &  0&2  & 0&34(6)   &$<0$&1     &162(8)   \\
F23    &  0&9  & 0&23(3)   &$<0$&11    &158(14)  \\
F24    &  2&9  & 0&19(1)   &$<0$&03    &165(3)   \\
\enddata
\tablenotetext{a}{the deconvolved angular size of the masers listed (major axis)$\times$(minor axis)}
\end{deluxetable}

\clearpage

\begin{deluxetable}{ l  r@{.}l r r}
\tablecaption{VLBA Maser Sizes and Flux Densities\label{vlbasizes}}
\tablecolumns{5}
\tablewidth{0pt}
\tablehead{
\colhead{Feature} & \multicolumn{2}{c}{$S$ (Jy)} & \colhead{angular size (mas)} & \colhead{P.A.\ (deg)}
}
\startdata
E11A    &   1&2  & 24(6)& 133(18) \\          
D18     &   0&5  & $<13$& 158(20) \\
F24     &   2&0  & 19(4)& 147(9) \\
\enddata 
\end{deluxetable}

\clearpage

\begin{deluxetable}{ l  r@{\ }r@{\ }l  r@{\ }r@{\ }l  c  r@{.}l  r@{.}l  r@{.}l }
\tablecaption{MERLIN Fitted Maser Properties\label{merpos}}
\tablecolumns{14}
\tablewidth{0pt}
\tablehead{
\colhead{Feature} & \multicolumn{3}{c}{R.A.\ (J2000)} & \multicolumn{3}{c}{Decl.\ (J2000)} & \colhead{$I$} & \multicolumn{2}{c}{$\Delta{v}$} & \multicolumn{2}{c}{$v_{\rm LSR}$} & \multicolumn{2}{c}{$T_B$} \\
 & \colhead{(h} & \colhead{m} & \colhead{s)} & \colhead{(\arcdeg} & \colhead{\arcmin} & \colhead{\arcsec)} & \colhead{(\mjb)} & \multicolumn{2}{c}{(\kms)} & \multicolumn{2}{c}{(\kms)} & \multicolumn{2}{c}{($10^6$~K)} \\
}
\startdata
B7     & 18&56&01.291  &    01&12&44.80  &   760  & 0&65(1) & 44&63(1)  &  11&   \\
C8A    &   &  &03.784  &      &08&46.97  &   340  & 0&93(2) & 46&16(1)  &   4&7  \\
C8B    &   &  &03.788  &      &  &46.51  &   230  & 0&87(3) & 46&23(1)  &   3&2  \\
E11A   &   &  &26.734  &      &29&18.15  &  2410  & 1&03(1) & 44&11(1)  &  34&   \\
E11B   &   &  &26.741  &      &  &16.59  &   200  & 0&84(3) & 44&23(1)  &   2&8  \\
E11C   &   &  &26.779  &      &  &17.53  &   360  & 0&71(1) & 44&15(1)  &   5&0  \\
E20A   &   &  &29.684  &      &  &08.16  &   940  & 0&99(1) & 44&29(1)  &  13&   \\
E20B   &   &  &29.692  &      &  &07.42  &   170  & 1&70(3) & 44&16(2)  &   2&4  \\
F23    &   &  &36.633  &      &26&29.82  &   840  & 0&88(2) & 46&72(1)  &  12&   \\
F24    &   &  &36.675  &      &  &37.14  &  2390  & 0&71(1) & 46&96(1)  &  33&   \\
\enddata
\end{deluxetable}

\clearpage

\begin{deluxetable}{ l  r@{\ }r@{\ }l  r@{\ }r@{\ }l  c  r@{.}l  r@{.}l  r@{.}l }
\tablecaption{VLBA Fitted Maser Properties\label{vlbapos}}
\tablecolumns{14}
\tablewidth{0pt}
\tablehead{
\colhead{Feature} & \multicolumn{3}{c}{R.A.\ (J2000)} & \multicolumn{3}{c}{Decl.\ (J2000)} & \colhead{$I$} & \multicolumn{2}{c}{$\Delta{v}$} & \multicolumn{2}{c}{$v_{\rm LSR}$} & \multicolumn{2}{c}{$T_B$} \\
 & \colhead{(h} & \colhead{m} & \colhead{s)} & \colhead{(\arcdeg} & \colhead{\arcmin} & \colhead{\arcsec)} & \colhead{(\mjb)} & \multicolumn{2}{c}{(\kms)} & \multicolumn{2}{c}{(\kms)} & \multicolumn{2}{c}{($10^9$~K)} \\
}
\startdata
E11A    & 18&56&26.7347 &    01&29&18.154 &   680  & 0&63(3)  & 43&85(1)  &   0&9  \\
D18     &   &  &29.4822 &      &20&26.556 &   450  & 0&57(5)  & 43&79(2)  &   0&6  \\
F24AIII &   &  &36.6741 &      &26&37.115 &   340  & 0&55(5)  & 46&84(2)  &   0&4  \\
F24AI   &   &  &36.6748 &      &  &37.136 &   970  & 0&59(3)  & 46&79(1)  &   1&3  \\
        &   &  &        &      &  &       &   140  & \multicolumn{2}{c}{\tablenotemark{a}} & 46&19(5)  &   0&2  \\
F24AII  &   &  &36.6761 &      &  &37.100 &   570  & 0&48(4)  & 46&77(2)  &   0&7  \\
        &   &  &        &      &  &       &   220  & \multicolumn{2}{c}{\tablenotemark{a}} & 46&34(3)  &   0&3  \\
\enddata
\tablenotetext{a}{insufficient signal to noise to fit line width}
\end{deluxetable}

\clearpage

\begin{deluxetable}{ l  r@{.}l  c  r@{.}l@{$\pm$}r@{.}l   r@{.}l@{$\pm$}r@{.}l }
\tablecaption{MERLIN Zeeman Results\label{circ}}
\tablecolumns{12}
\tablewidth{0pt}
\tablehead{
\colhead{Feature} & \multicolumn{2}{c}{$v_{\rm RCP} - v_{\rm LCP}$\tablenotemark{a}} & \colhead{$x_B$} & \multicolumn{4}{c}{$B_{\rm RL}$} & \multicolumn{4}{c}{$B_{\theta}$\tablenotemark{b}} \\
      & \multicolumn{2}{c}{(km/s)} &  & \multicolumn{4}{c}{(mG)} & \multicolumn{4}{c}{(mG)}
}
\startdata
B7      & \multicolumn{2}{c}{\tablenotemark{c}} & \tablenotemark{c} & \multicolumn{4}{c}{\tablenotemark{c}} & $-0$&33 & 0&09 \\
C8A     & \multicolumn{2}{c}{\tablenotemark{c}} & \tablenotemark{c} & \multicolumn{4}{c}{\tablenotemark{c}} & \multicolumn{4}{c}{\tablenotemark{d}} \\
C8B     & \multicolumn{2}{c}{\tablenotemark{c}} & \tablenotemark{c} & \multicolumn{4}{c}{\tablenotemark{c}} & \multicolumn{4}{c}{\tablenotemark{d}} \\
E11A    & $-0$&084(4)  & 0.08 & $-0$&73 & 0&02 & $-0$&75 & 0&02 \\
E11B    & \multicolumn{2}{c}{\tablenotemark{c}} & \tablenotemark{c} & \multicolumn{4}{c}{\tablenotemark{c}} & \multicolumn{4}{c}{\tablenotemark{d}} \\
E11C    & $-0$&066(13) & 0.09 & $-0$&58 & 0&09 & $-0$&57 & 0&08 \\
E20A    & $-0$&049(7)  & 0.05 & $-0$&43 & 0&07 & $-0$&48 & 0&02 \\
E20B    & \multicolumn{2}{c}{\tablenotemark{c}} & \tablenotemark{c} & \multicolumn{4}{c}{\tablenotemark{c}} & \multicolumn{4}{c}{\tablenotemark{d}} \\
F23     & $-0$&061(11) & 0.08 & $-0$&53 & 0&09 & $-0$&53 & 0&03 \\
F24     & $-0$&061(4)  & 0.09 & $-0$&53 & 0&05 & $-0$&56 & 0&02 \\
\enddata
\tablenotetext{a}{the velocity difference between the centers of the Gaussians fitted to the RCP and LCP line profiles}
\tablenotetext{b}{the magnetic field strength fitted to $V/(\partial{I}/\partial{\nu})$}
\tablenotetext{c}{insufficient signal to noise to resolve Zeeman splitting between RCP and LCP}
\tablenotetext{d}{insufficient signal to noise to observe Zeeman pattern in Stokes $V$}
\end{deluxetable}

\clearpage

\begin{deluxetable}{ l  c  c  c  r  c }
\tablecaption{MERLIN Linear Polarization Results\label{merpol}}
\tablecolumns{5}
\tablewidth{0pt}
\tablehead{
\colhead{Feature} & \colhead{$Q$} & \colhead{$U$} & \colhead{$\chi$} & \colhead{$q$} \\
 & \colhead{(\mjb)} & \colhead{(\mjb)} & \colhead{(deg)} & \colhead{(\%)} \\
}
\startdata
B7     & \tablenotemark{a} & \tablenotemark{a} & \tablenotemark{a} & \tablenotemark{a} \\
C8A    & \tablenotemark{a} & \tablenotemark{a} & \tablenotemark{a} & \tablenotemark{a} \\
C8B    & \tablenotemark{a} & \tablenotemark{a} & \tablenotemark{a} & \tablenotemark{a} \\
E11A   &   $200$ & $-220$ & $-24$(1)  & 12(1) \\
E11B   &   $<10$ &  $<10$ &    -      & $<9$  \\
E11C   &   $-15$ &  $-50$ & $-53$(6)  & 14(3) \\
E20A   &   $ 25$ &  $-30$ & $-26$(7)  &  4(1) \\
E20B   &   $<10$ &  $<10$ &    -      & $<8$  \\
F23    &   $-60$ &  $<10$ & $  0$(4)  &  7(3) \\
F24    &   $ 70$ & $-150$ & $-32$(3)  &  7(1) \\
\enddata
\tablenotetext{a}{polarization calibration unavailable}
\end{deluxetable}


\begin{thebibliography}{}
\bibitem[Abramowitz \& Stegun (1972)]{abr72} Abramowitz, M.\ \& Stegun, I.\ A.\ (Eds.) 1972 Handbook of Mathematical Functions with Formulas, Graphs, and Mathematical Tables (New York: Dover)
\bibitem[Arikawa et al.\ (1999)]{ari99} Arikawa, Y., Tatematsu, K., Sekimoto, Y., \& Takahashi, T.\ 1999, \pasj, 51, L7
\bibitem[Bertoldi \& McKee (1992)]{ber92} Bertoldi, F. \& McKee, C.\ F.\ 1992, \apj, 395, 140
\bibitem[Ball \& Staelin (1968)]{ball68} Ball J.\ A.\ \& Staelin, D.\ H.\ 1968, \apjl, 153, L41
\bibitem[Balsara et al.\ (2001)]{bal01} Balsara, Dinshaw, Benjamin, Robert A., \& Cox, Donald P.\ 2001, \apj, 563, 800
\bibitem[Briggs et al.\ (1999)]{bri99} Briggs, Daniel S., Schwab, Frederic R., \& Sramek, Richard, A.\ 1999, in Synthesis Imaging in Radio Astronomy II. Edited by G.\ B.\ Taylor, C.\ L.\ Carilli, and R.\ A.\ Perley. ASP Conference Series, 180, 127
\bibitem[Brogan et al.\ (2000)]{bro00} Brogan, C.\ L., Frail, D.\ A., Goss, W.\ M., \& Troland, T.\ H.\ 2000, \apj, 537, 875 (BFGT)
\bibitem[Caswell et al.\ (1975)]{cas75} Caswell, J.\ L., Murray, J.\ D., Roger, R.\ S., Cole, D.\ J., \& Cooke, D.\ J.\ 1975, \aap, 45, 239
\bibitem[C97]{C97} Claussen, M.\ J, Frail, D.\ A., Goss, W.\ M., \& Gaume, R.\ A.\ 1997, \apj, 489, 143 (C97)
\bibitem[C99]{C99} Claussen, M.\ J, Goss, W.\ M., Frail, D.\ A., \& Desai, K.\ 1999, \apj, 522, 349 (C99)
\bibitem[Claussen et al.\ (2002)]{cla02} Claussen, M.\ J, Goss, W.\ M., Desai, K.\ M., \& Brogan, C.\ L.\ 2002, \apj, 580, 909
\bibitem[Chevalier (1999)]{che99} Chevalier, Roger A.\ 1999, \apj, 511, 798
\bibitem[Davies (1974)]{dav74} Davies, R.\ D.\ 1974, in {\it IAU Symposium 60, Galactic Radio Astronomy}, ed.\ F.\ J.\ Kerr \& S.\ C.\ Simonson, III (Dordrecht: Reidel) p.\ 275
\bibitem[Dickel \& Milne (1976)]{dic76} Dickel, John R.\ \& Milne, D.\ K.\ 1976, Aust.J.Ph., 29, 435
\bibitem[Dubner et al.\ (2004)]{dub04} Dubner, G., Giacani, E., Reynoso, E., \& Par\'{o}n, S.\ 2004, \aap, 426, 201
\bibitem[Elitzur (1976)]{eli76} Elitzur, M.\ 1976, \apj, 203, 124
\bibitem[Elitzur (1992)]{eli92} Elitzur, M.\ 1992 Astronomical Masers (Dordrecht:Kluwer)
\bibitem[Elitzur (1996)]{eli96} Elitzur, M.\ 1996, \apj, 457, 415
\bibitem[Elitzur (1998)]{eli98} Elitzur, M.\ 1998, \apj, 504, 390
\bibitem[Frail et al.\ (1994)]{fra94} Frail, D.\ A., Goss, W.\ M., \& Slysh, V.\ I.\ 1994, \apjl, 424, L111
\bibitem[Frail et al.\ (1996)]{fra96} Frail, D.\ A., Goss, W.\ M., Reynoso, E.\ M., Giacani, E.\ B., Green, A.\ J., \& Otrupcek, R.\ 1996, \aj, 111, 1651
\bibitem[Frail \& Mitchell (1998)]{fra98} Frail, Dale A.\ \& Mitchell, George F.\ 1998, \apj, 508, 690
\bibitem[Georgelin \& Georgelin (1976)]{geo76} Georgelin, Y.\ M.\ \& Georgelin, Y.\ P.\ 1976, \aap, 49, 57
\bibitem[Giacani et al.\ (1997)]{gia97} Giacani, E.\ B., Dubner, G.\ M., Kassim, N.\ E., Frail, D.\ A., Goss, W.\ M., Winkler, P.\ F., \& Williams, B.\ F.\ 1997, \aj, 113, 1379
\bibitem[Gray (2003)]{gra03} Gray, M.\ D.\ 2003, \mnras, 343, L33
\bibitem[Green et al.\ (1997)]{gre97} Green, A.\ J., Frail, D.\ A., Goss, W.\ M., \& Otrupcek, R.\ 1997, \aj, 114, 2058
\bibitem[Goss (1968)]{gos68} Goss, W.\ M.\ 1968, \apjs, 15, 31
\bibitem[Goss \& Robinson (1968)]{gr68} Goss, W.\ M.\ \& Robinson, B.\ J.\ 1968, \aplett, 2, 81
\bibitem[Goss et al.\ (1971)]{gos71} Goss, W.\ M., Caswell, J.\ L., \& Robinson B.\ J.\ 1971, \aap, 14, 481
\bibitem[Hoffman et al.\ (2003a)]{H03} Hoffman, I.\ M., Goss, W.\ M., Brogan, C.\ L., Claussen, M.\ J, \& Richards, A.\ M.\ S.\ 2003a, \apj, 583, 272 (H03)
\bibitem[Hoffman et al.\ (2003b)]{hof03} Hoffman, Ian M., Goss, W.\ M., Palmer, Patrick, \& Richards, A.\ M.\ S.\ 2003b, \apj, 598, 1061
\bibitem[Hoffman et al.\ (2005)]{hof05} Hoffman, Ian M., Goss, W.\ M., Brogan, C.\ L., Claussen, M.\ J 2005, \apj, 620, 257 (Paper I)
\bibitem[Kodaira et al.\ (1977)]{kod77} Kodaira, S., Ishii, K., Nakamura, T., Inatani, J., Nagane, K., Tojo, A., \& Sato, F.\ 1977, \pasj, 29, 53
\bibitem[Koo \& Heiles (1995)]{koo95} Koo, B.-C.\ \& Heiles, C.\ 1995, \apj, 442, 679
\bibitem[Koralesky et al.\ (1998)]{kor98} Koralesky, B., Frail, D.\ A., Goss, W.\ M., Claussen, M.\ J, \& Green, A.\ J.\ 1998, \aj, 116, 1323
\bibitem[Kundu \& Velusamy (1972)]{kun72} Kundu, M.\ R.\ \& Velusamy, T.\ 1972, \aap, 20, 237
\bibitem[Lazendic et al.\ (2004)]{laz04} Lazendic, J.\ S., Wardle, M., Burton, M.\ G., Yusef-Zadeh, F., Green, A.\ J., \& Whiteoak, J.\ B.\ 2004, \mnras, 354, 393
\bibitem[LGE]{LGE} Lockett, P., Gauthier, E., \& Elitzur, M.\ 1999, \apj, 511, 235
\bibitem[Milne \& Dickel (1975)]{mil75} Milne, D.\ K.\ \& Dickel, J.\ R.\ 1975, Aust.J.Ph., 28, 209
\bibitem[Milne (1987)]{mil87} Milne, D.\ K.\ 1987, Aust.J.Ph., 40, 771
\bibitem[Palmer et al.\ (2003)]{pal03} Palmer, Patrick, Goss, W.\ M., \& Devine, K.\ E.\ 2003, \apj, 599, 324
\bibitem[Pavlakis \& Kylafis (1996)]{pav96} Pavlakis, Konstantinos G.\ \& Kylafis, Nikolaos D.\ 1996, \apj, 467, 309
\bibitem[Press et al.\ (2001)]{pre01} Press, William H., Teukolsky, Saul A., Vetterling, William T., \& Flannery, Brian P.\ 2001 Numerical Recipes in Fortran 77 (Cambridge: Cambridge University Press)
\bibitem[Radhakrishnan et al.\ (1972)]{rad72} Radhakrishnan, V., Goss, W.\ M., Murray, J.\ D., \& Brooks, J.\ W.\ 1972, \apjs, 24, 49
\bibitem[Reynoso \& Mangum (2000)]{rey00} Reynoso, Estela M.\ \& Mangum, Jeffrey G.\ 2000, \apj, 545, 874
\bibitem[Roberts et al.\ (1993)]{rob93} Roberts, D.\ A., Crutcher, R.\ M., Troland, T.\ H., \& Goss, W.\ M.\ 1993, \apj, 412, 675
\bibitem[Robinson et al.\ (1970)]{rob70} Robinson, B.\ J., Goss, W.\ M., \& Manchester, R.\ N.\ 1970, Aust.J.Ph., 23, 363
\bibitem[Rybicki \& Lightman (1979)]{ryb79} Rybicki, George B.\ \& Lightman, Alan, P.\ 1979 Radiative Processes in Astrophysics (Wiley: New York)
\bibitem[Sarma et al.\ (2001)]{sar01} Sarma, A.\ P., Troland, T.\ H., \& Romney, J.\ D.\ 2001, \apjl, 554, L217
\bibitem[Sato (1986)]{sat86} Sato, F.\ 1986, \aj, 91, 378
\bibitem[Seta et al.\ (2004)]{set04} Seta, Masumichi et al.\ 2004, \aj, 127, 1098
\bibitem[Turner (1969)]{tur69} Turner, B.\ E.\ 1969, \aj, 74, 985
\bibitem[Vlemmings et al.\ (2002)]{vle02} Vlemmings, W.\ H.\ T., Diamond, P.\ J., \& van Langevelde, H.\ J.\ 2002, \aap, 394, 589
\bibitem[Wardle (1999)]{war99} Wardle, M. 1999, \apjl, 505, L101
\bibitem[Watson \& Wyld (2001)]{wat01} Watson, W.\ D.\ \& Wyld, H.\ W.\ 2001, \apjl, 558, L55 (WW01)
\bibitem[Watson et al.\ (2002)]{wat02} Watson, W.\ D., Sarma, A.\ P., \& Singleton, M.\ S.\ 2002, \apjl, 570, L37
\bibitem[Wolszczan et al.\ (1991)]{wol91} Wolszczan, A., Cordes, J.\ M., \& Dewey, R.\ J.\ 1991, \apjl, 372, L99
\bibitem[Wootten (1977)]{woo77} Wootten, H.\ A.\ 1977, \apj, 216, 440
\bibitem[Yusef-Zadeh et al.\ (1996)]{yus96} Yusef-Zadeh, F., Roberts, D.\ A., Goss, W.\ M., Frail, D.\ A., \& Green, A.\ J.\ 1996, \apjl, 466, L25
\bibitem[Yusef-Zadeh et al.\ (1999)]{yus99a} Yusef-Zadeh, F., Roberts, D.\ A., Goss, W.\ M., Frail, D.\ A., \& Green, A.\ J.\ 1999a, \apj, 512, 230
\bibitem[Yusef-Zadeh et al.\ (1999)]{yus99b} Yusef-Zadeh, F., Goss, W.\ M., Roberts, D.\ A., Robinson, B., \& Frail, D.\ A.\ 1999b, \apj, 527, 172
\bibitem[Yusef-Zadeh et al.\ (2000)]{yus00} Yusef-Zadeh, F., Shure, M., Wardle, M., \& Kassim, N.\ 2000, \apj, 540, 842
\end{thebibliography}
\end{document}